\documentclass[a4paper,usenatbib]{pasn}

\usepackage[T1]{fontenc}
\usepackage{ae,aecompl}

\usepackage{graphicx}	
\usepackage{amsmath}	
\usepackage{amssymb}	

\title[Cosmic Ray Diurnal Anisotropy and Forbush Event Simultaneity]{The Implication of Enhanced Cosmic Ray Diurnal Anisotropy on the Global Simultaneity of Forbush Decreases: A Statistical Approach}
\author[Okike et al.]{
O. Okike$^{1,4}$\thanks{Email:giftedlife2014@gmail.com}, 
J. A. Alhassan$^{2,4}$
and
I. O. Eya$^{2,3,4}$,
\\
$^{1}$Department of Industrial Physics, Ebonyi State University, Abakaliki, Nigeria.\\
$^{2}$Department of Physics and Astronomy, University of Nigeria, Nsukka, Nigeria.\\
$^{3}$Physics/Electronics Technique -- Department of Science Laboratory Technology, University of Nigeria, Nsukka, Nigeria.\\
$^{4}$Astronomy and Astrophysics Lab. Faculty of Physical Sciences, University of Nigeria, Nsukka.
}
\date{Accepted November 05, 2022, Received October 02, 2022}

\pubyear{2022}

\begin{document}
\label{firstpage}
\pagerange{\pageref{firstpage}--\pageref{lastpage}}
\setcounter{page}{44}
\maketitle

\begin{abstract}
The short-term rapid CR flux depressions, generally referred to as Forbush decreases (FDs), are the most spectacular time-intensity CR variation. The need for analytical transformation of the observational CR time series data, to account for FDs and other recurrence tendencies such as periodicities and cycles, was noted since the 1930s'. Nevertheless, it has been recently observed that harmonic analysis, which is capable of transforming raw CR data into different frequencies, is rarely exploited. Predominant in the literature are the ordinary Fourier and power spectral analyses, generally used to calculate the positive vectors (amplitude and phase) of the periodic diurnal CR anisotropy. In the two approaches, the days immediately connected with FDs are frequently removed to minimize unusual changes in the amplitude of the vectors as well as a spurious time of maximum. However, there is a paucity of publications that adjust for the influence of enhanced CR diurnal anisotropy on the magnitude and timing of FDs. Recently, in an attempt to test the global FD event simultaneity, a combination of numerical filtering and fast Fourier transform techniques was deployed to account for these superposition tendencies in daily CR data, including the intractable CR diurnal anisotropy. However, an extremely sensitive version of the software would be required to analyze high-resolution CR hourly averages. As a way of achieving accurate detection and precise timing of FD signals, the computer algorithms were technically improved and employed in a long-term statistical investigation. To appreciate the implemented extremely sensitive statistical technique, several validation analyses, including FD-based solar-terrestrial correlation, comparison of FD catalogues, and Frobush event simultaneity test were conducted.
\end{abstract}
\begin{keywords}
Cosmic rays, cosmic ray modulation, Forbush decreases, algorithm, simultaneity, diurnal anisotropy
\end{keywords}



\section{Introduction}\label{sec1}
Investigation of temporal changes in the galactic cosmic ray (GCR) radiation
at the Earth started in the 1930s following the pioneering work of S. E. Forbush
\citep{1}. Among the various CR intensity-time changes such as daily, 27-day, 11-
and 22-year periodic cycles, the sharp and non-recurring (although they may
be recurrent when connected with coronal holes) intensity decreases, the so-
called Forbush decreases (FDs), which seem to have gained more attention. \cite{2} is an
excellent review of the literature on FDs covering the initial understanding
of Forbush events from 1938-1970 whereas \citep{3,4,5} are more focused on causes
of the phenomenon. Several interesting FD-related subjects (e.g. anisotropies
as well as their attendant bias on FD magnitude estimation, magnitudes of
FDs, CR modulation during FDs, FD dependence on rigidity and altitude, the
the link between FD and geomagnetic disturbances, longitudinal and latitudinal
dependence of FDs, FD onset time, precursory increases/decreases that occur
before the arrival of the solar events that generate FDs, the recovery of intensity
level during FDs and simultaneity of FDs), was reviewed by \cite{2}. 

Almost each of these topics is still hotly discussed among scientists, especially in connection with Sun-weather studies \citep[e.g][]{6,7,8,9,10,11,12,13,14,15,16,17,18,19,20}. For example, \cite{20} investigated solar-terrestrial relations concerning FD onset time/time of maximal decreases using the superposed epoch method of analysis. Besides the customary superposition analysis, correlation and regression analyses have also been used to test the connection between FDs and solar-geophysical parameters \citep[see][]{18,19,21,22,23,24}. Despite the volumes of work on the influence of the Sun (using FD as an indicator) on the Earth's weather, technology, and human life, definitive conclusions have yet to be reached. A review of the existing publications suggests that the definition of FDs, causes of FDs, Forbush event identification and timing, the implication of CR anisotropy on FD event characteristics, simultaneity and magnitude of Forbush events as well as what determines them are some of the major sources of misunderstanding among CR researchers \citep{21,25,26,27,28,29,30,31,32,33,34}. In the current work, we intend to further investigate the phenomena of global simultaneity of Forbush events. 

\section{Data}
The hourly CR data are sourced from \url{http://   cr0.izmiran.rssi.ru/} while data on IZMIRAN FEs are taken from\\
\url{http:// spaceweather. izmiran. ru/ eng/dbs.html}. 
The latter will be referred to as the Forbush-effects and interplanetary disturbances database (FEID, hereafter.) Solar wind speed (SWS) and interplanetary magnetic field (IMF) data used are taken from \url{https://omniweb.gsfc.nasa.gov/html/ow_data.html}. Accessing SWS and IMF data is straightforward. The principal investigators provide a simple interface that helps researchers produce plots or download text files of the selected data resolutions. Hourly, daily, 27-day, and yearly averaged data are available on the website. There is equally a window for data availability checks. The hourly data used in the current work were downloaded by filling out the selection icons for SWP and IMF as well as the submission box.

\section{Method of Analysis}
The manual method of event identification is considered a case study approach. Specific events that happen within the period of interest are isolated and examined in detail. How much selection takes place remains an open question in such method \citep{35}. A statistical approach to FD event identification is attempted here.

Figure \ref{Figure3} shows the CR flux variation for a full solar cycle (Solar Circle 23) for THUL and INVK stations. The dips are indications of FDs (see \citep{36}. There are several of them. We have marked some of the large events for reference purposes. It is also evident that the largest FDs happen around the solar cycle maximum (2001). Instead of isolating each of the pits and separately calculating their magnitude, onset time, and time of minimum, the programs developed by \cite{31}(Paper I, hereafter) and \cite{45} will be enhanced and used to calculate the event magnitude and time of occurrence of all the FDs that happen within the period.

Detection of FDs from hourly data requires a relatively more efficient and sensitive algorithm than those used in the analysis of daily data. It is interesting to note that two versions of \cite{45} codes $-$ which takes CR hourly means as input data $-$ are recently developed and implemented in \cite{37} and \cite{38}. Using one and five years' data respectively, \cite{37} and \cite{38} demonstrated that different versions of our FD algorithm can detect FDs from hourly CR counts. An attempt is made to briefly describe the codes in the next section.

\subsection{A Brief Description of the FD Location Algorithms}
\cite{45} employed two different FD selection programs. The first code (referred to as Prog1, hereafter) was developed by a group of researchers \citep[see][]{31}. The second program (hereafter, Prog2) was recently developed and first implemented by \cite{45}. 

\begin{center}
	\begin{figure}
	\centering
	\includegraphics[width=10cm]{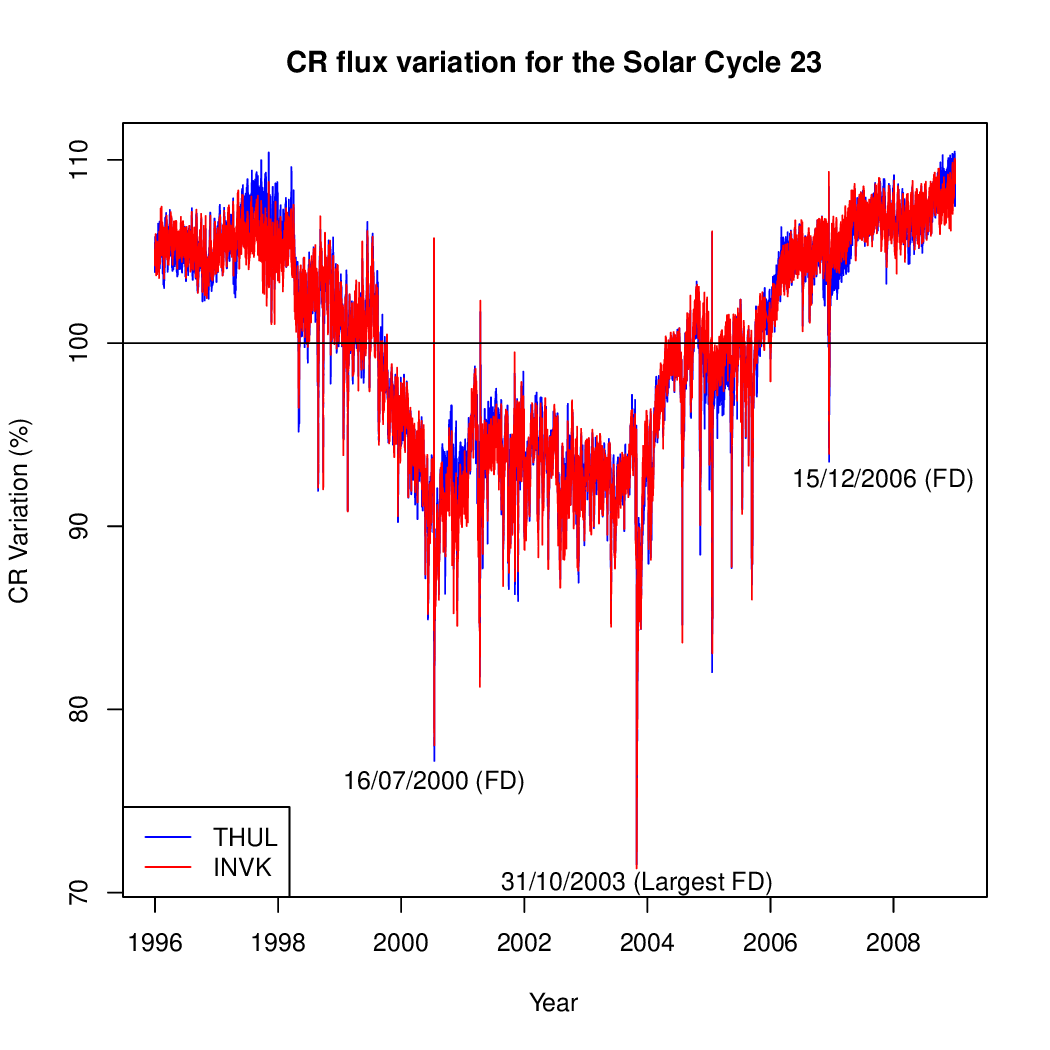}
	\caption{Raw CR intensity variation for THUL (blue) and INVK (red) stations for the Solar Cycle 23}
	\label{Figure3}
	 \end{figure}
	\end{center}

Prog1 employs a combination of the traditional filtering method \citep{36,39,40} and fast Fourier transform techniques. The two major signals that could complicate the identification of the rapid and non-periodic fluctuations (e.g. FDs and GLEs) in CR raw data are solar cycle variations and the periodic as well as the enhanced diurnal variations. The effect of the 11-year solar cycle is evident in Figure \ref{Figure3}. While it is easy to filter the solar cycle effect using a high pass filter, adjusting for the contribution of CR diurnal anisotropy requires some data transformation. Before passing the raw data to the Fourier transform subroutines, the data values were first subtracted from the running mean to remove the effect of solar cycle variations. The running mean is calculated using a special function (roll mean) in R for statistical computing software \citep{41}.

After removing the solar cycle effects, the next challenge is how to deal with the superposed ordered/periodic and random/non-periodic components of the data. The Fourier transform code handles this stage. It separates the data into two components $-$ regular and irregular frequency parts. The regular frequency part accounts for the diurnal variation whereas the irregular component is composed of the short-term random intensity changes, including FDs and GLEs.

The two separate components are of interest in the current work. Amplitudes of FDs and diurnal waves are calculated respectively from the irregular and ordered frequency signals. Although this could be done manually, it will be tedious and time-consuming. Another program capable of simultaneously tracking event time and magnitude is designed to do the calculation. The algorithm scans input signals for possible maxima/peaks and dips/pits/minima. The peaks (note that most of the peaks have been removed in Figure \ref{Figure3} for plotting purposes) may be GLEs or anomalous CR modulations/enhancement \citep{42}. Since we are interested in the depressions, the program is instructed to calculate the time and magnitude of the intensity reductions. 

Prog2 operates basically in the same way as Prog1. There are, however, a few similarities as well as differences in the operation of Prog2 and Prog1. The high pass filter and rollmean functions are integrated into both software. FD location subroutines are also implemented in both programs. The major difference between Prog1 and Prog2 is that the influence of anisotropy is not taken into consideration in Prog2. 

We note that the date/time format implemented by \cite{45} is of the form "YYYY.mm.dd" whereas the format in the current work is of the form "YYYY.mm.dd HH: mm". This is to enable both Prog1 and Prog2 to take hourly data as its input signal.

\subsection{Application of Prog1 and Prog2}

\subsubsection{Low Solar Activity Period}

\begin{center}
	\begin{figure}
	\centering
	\includegraphics[width=10cm]{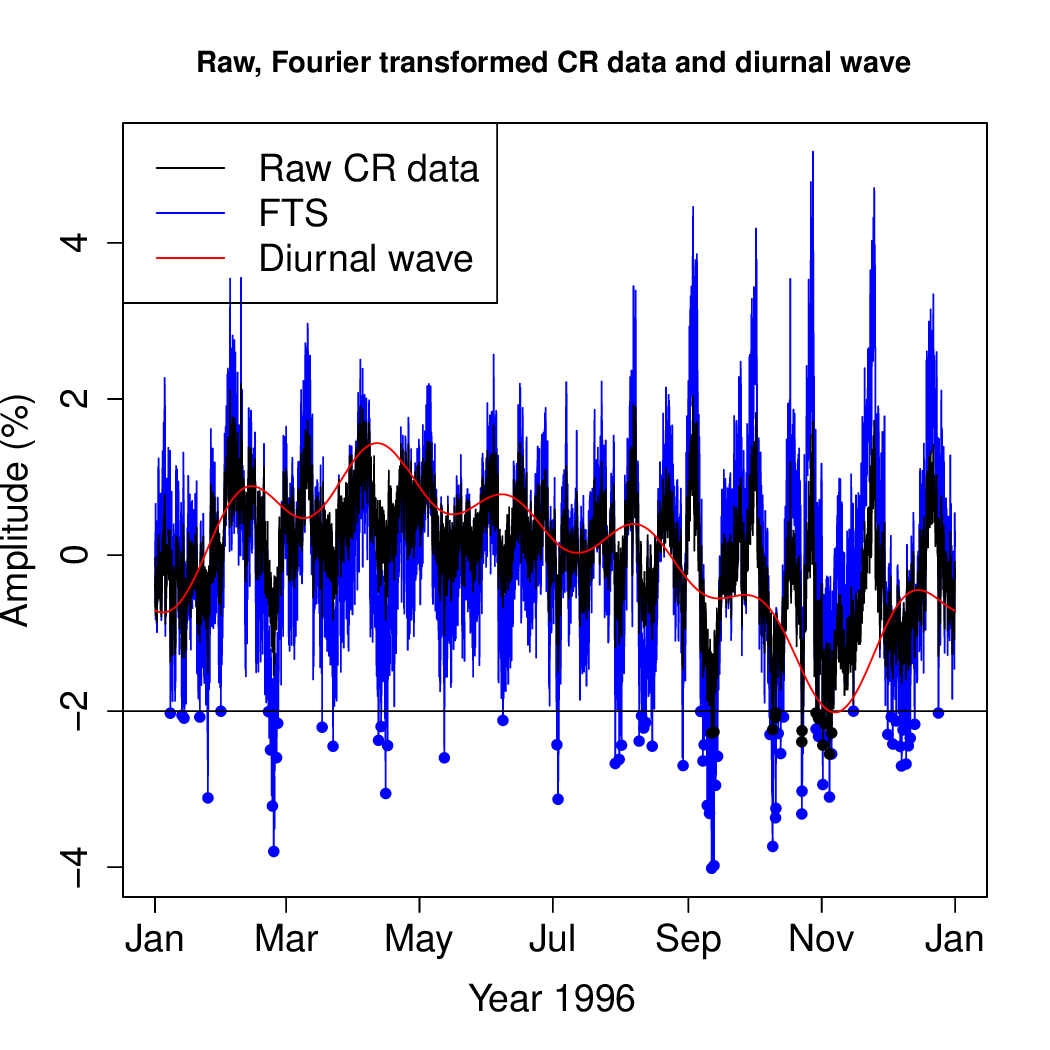}
	\caption{Comparison of raw, Fourier transformed (FTS) and diurnal wave at INVK station for the year 1996. The filled blue and black circles are indications of FDs associated with the (FTS) and the raw data.}
	\label{Figure4}
	 \end{figure}
	\end{center}

\begin{center}
	\begin{figure}
	\centering
	\includegraphics[width=10cm]{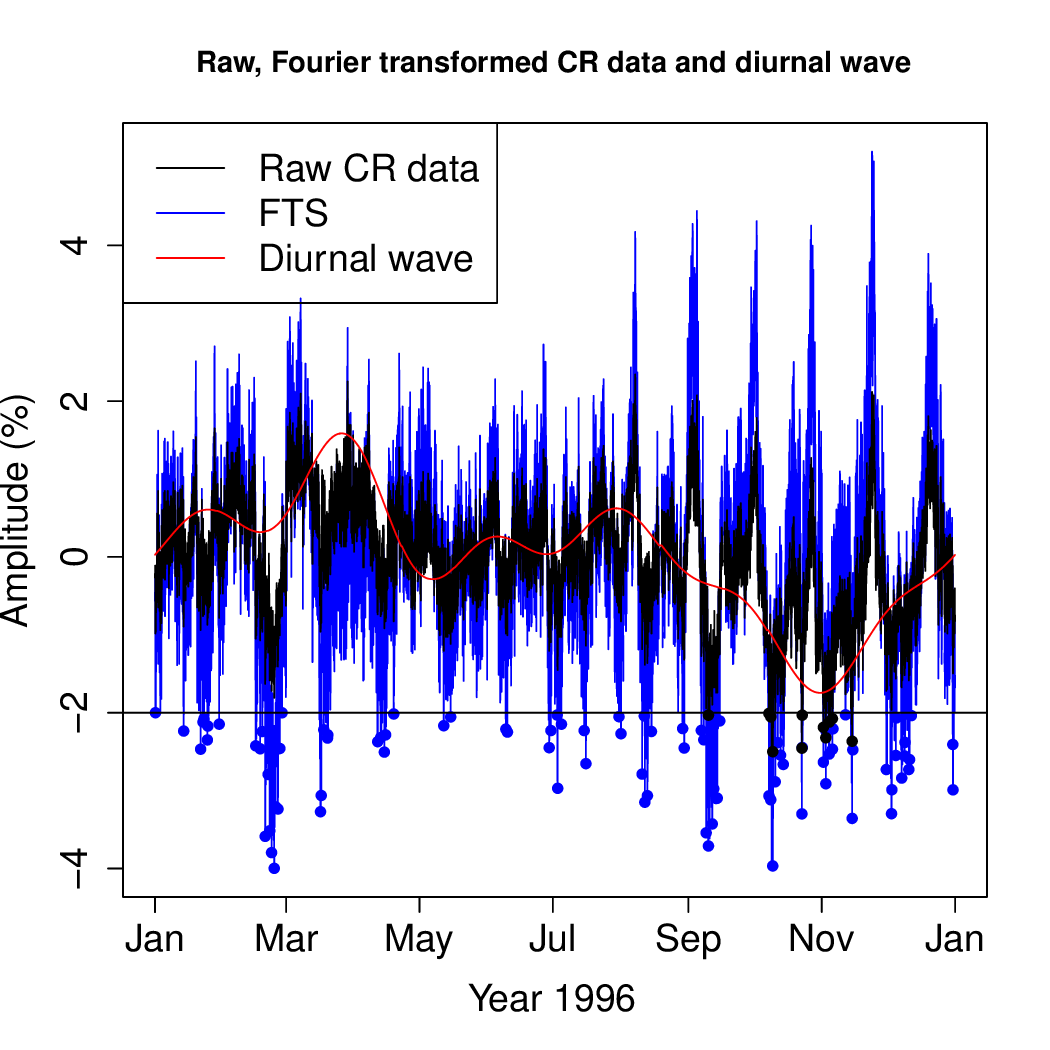}
	\caption{Same as Figure \ref{Figure4} for THUL station.}
	\label{Figure5}
	 \end{figure}
	\end{center}  

This section aims to demonstrate the performance of the two FD detection algorithms. For clarity's sake, two separate years (1996 and 2003) are used to demonstrate the operations of Prog1 and Prog2. The two years represent the periods of low (1996) and 2003 (high) solar activities. Figures \ref{Figure4} and \ref{Figure5} show the variation patterns of the raw data, the Fourier transformed component (FTS), and the diurnal wave at THUL and INVK stations. The blue and black filed circles are FDs detected by Prog1 and Prog2. It is evident from Figure \ref{Figure3} that CR intensity reductions in 1996 are very small compared to the period of high solar activity. Manually selecting FDs from the raw data might be very difficult. Inspection of Figures \ref{Figure4} and \ref{Figure5} shows that the FDs associated with the raw data are fewer (30 and 21 at INVK and THUL respectively) in comparison with those detected from the FTS (69 and 90 at INVK and THUL respectively). A small threshold (CR (\%) $\leq -2$) was employed here.

Whereas other researchers \citep[e.g][]{43} that searched for FDs within the Solar Cycle 23 did not identify FDs in the year 1996, the IZMIRAN group recorded 97 FDs the same year. This is possible with the aid of the GSM which accounts for CR anisotropy. The red curve in the two diagrams is representative of diurnal anisotropy at INVK and THUL stations. Separating them from the raw data allows us to detect the FDs that are otherwise, obscured. There are interesting differences and similarities in the trend of the diurnal anisotropy at the two stations. The largest amplitude of the anisotropy at INVK and THUL are respectively -2.02\% and -1.74\%. It is, however, better to present similar diagrams for a period of high solar activity for comparative purposes.

\begin{center}
	\begin{figure}
	\centering
	\includegraphics[width=10cm]{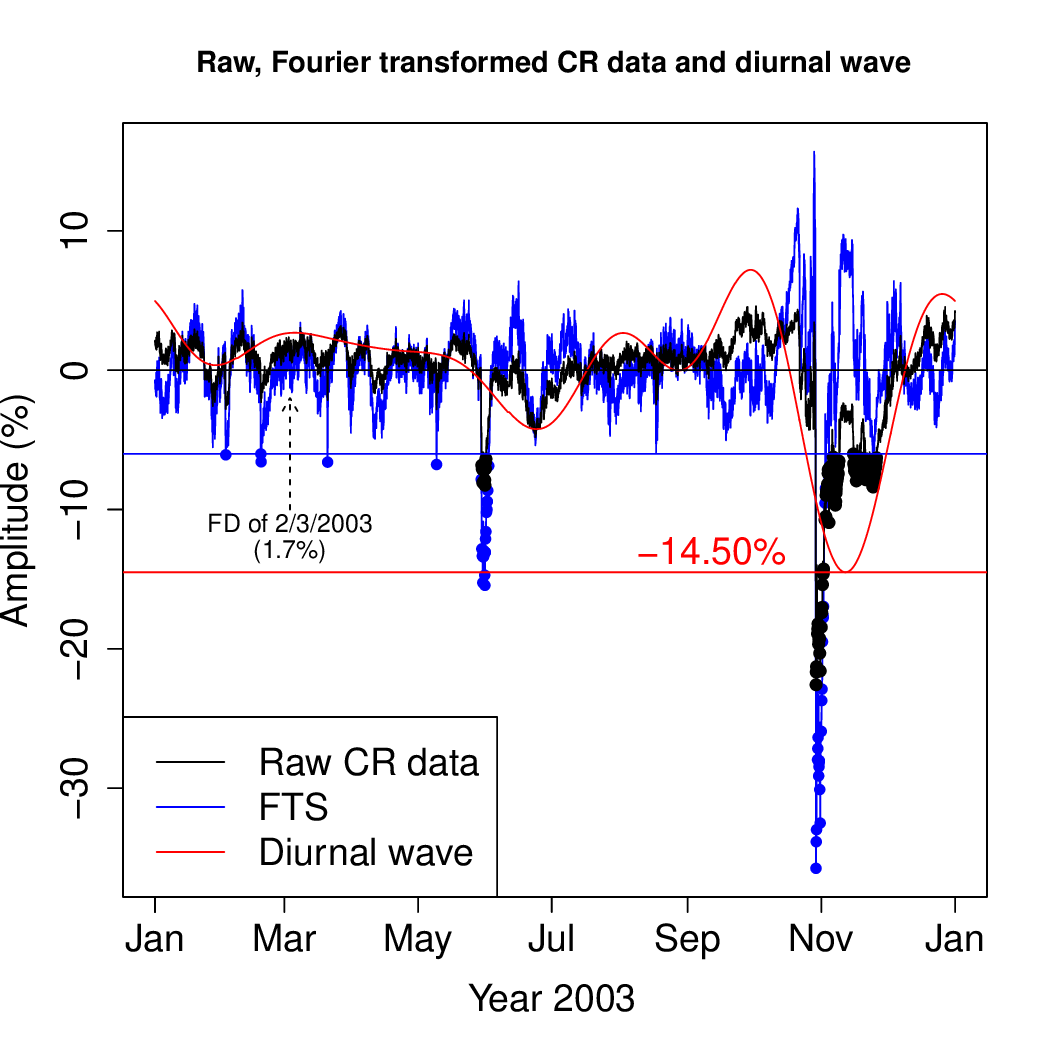}
	\caption{Same as Figure \ref{Figure4} for the year 2003.}
	\label{Figure6}
	 \end{figure}
	\end{center}  

\begin{center}
	\begin{figure}
	\centering
	\includegraphics[width=10cm]{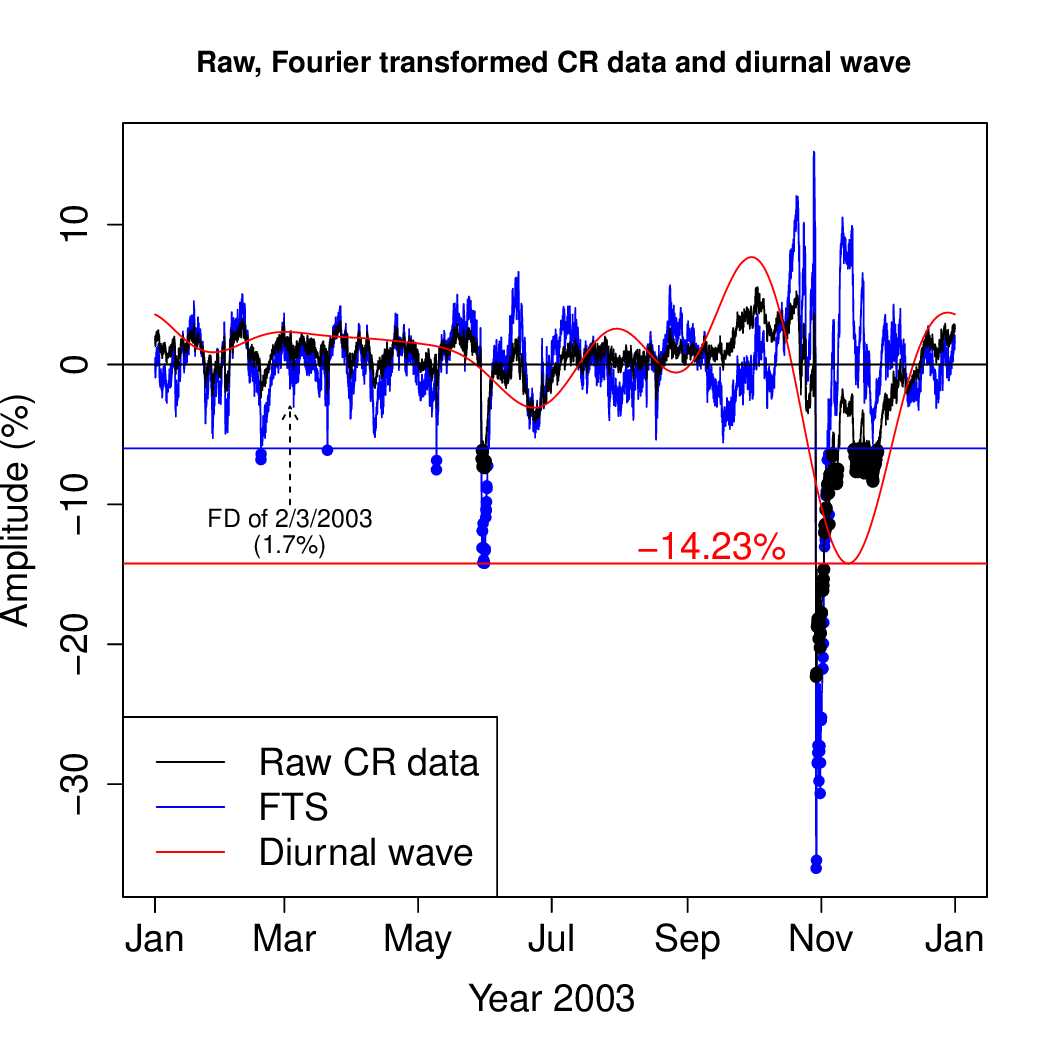}
	\caption{Same as Figure \ref{Figure5} for the year 2003.}
	\label{Figure7}
	 \end{figure}
	\end{center}  

\subsubsection{High Solar Activity Period}
Figures \ref{Figure6} and \ref{Figure7} present the relationship for the raw data, FTS, and diurnal wave for the year 2003 for INVK and THUL stations respectively. The blue and red horizontal lines indicate the selection baseline and the amplitude of the largest anisotropy that happened toward the end of 2003. As noted in Figures \ref{Figure4} and \ref{Figure5}, there are noticeable differences between the raw data and FTS. For example, the FTS indicates larger depressions than the raw data. The position of the small event of 2/3/2003 is marked in both diagrams. The arrow shows a small dip pointing below the horizontal black line for the FTS whereas a similar dip for the raw data is above the line in Figure \ref{Figure6}. We have a somewhat different scenario in Figure \ref{Figure7}. The corresponding depression in the raw data is below the zero line. The implications are that the event may not be detected from the raw data. But It may be detected from the raw data and the FTS in Figure \ref{Figure7}. 

However, whether Prog1 detects the event or not depends on the detection threshold. It is evident in both diagrams that the small event is not picked by the algorithms. This is due to the large normalization baseline (CR (\%) $\leq -6$) used here. Several dips larger than those that appear within the time of the event of 2/3/2003 are also not detected. When the threshold was increased to (CR (\%) $\leq -0.001$) Prog1 detected the event at 14:00 on 3/3/2003 at the INVK and THUL stations. Its magnitude is respectively -1.3 and -1.73\%. The event seems to be completely obscured by anisotropy and thus, it is not detected from the raw data. The small dip around the event in Figure \ref{Figure7} was another event that happened at 06:00 on 5/3/2003. Its magnitude is -0.36\%. 

The intensity variation trend of the diurnal wave in comparison with the raw and FTS is also quite interesting. There are two large depressions in Figures \ref{Figure6} and \ref{Figure7}. The two dips are accompanied by two large amplitudes of the diurnal wave. The largest FDs attracted the largest amplitudes of anisotropy in the two diagrams. This is not, however, the case in the period of low solar activity. The large amplitudes of the diurnal anisotropy in Figures \ref{Figure4} and \ref{Figure5} seem not to track the large FDs.

\subsection{Coincident and Non-Coincident Algorithms}
Event simultaneity could be defined regarding the complete FD profile or either of the parts of Forbush events such as the onset, main phase, FD minimum, and the recovery phase at different CR stations as was the case in \cite{25} and \citep{27}. In the light of \cite{45}, an event is considered to be simultaneous in the current work if its time of minimum happens at the same universal time at INVK and THUL CR stations. Conversely, an event is regarded as non-simultaneous if its FD minimum time differs at the two stations.

A simple coincident or matching code was developed by \citep{44}. The algorithm searches for equivalent FD magnitude or identical time in two or more event catalogs. The magnitude and event date of all the FDs from separate stations are first transformed into a data frame or matrix of equal or unequal length. The data are then passed to the algorithm, specifying the magnitude or event date/time as the key search term. For the timing of globally simultaneous or non-simultaneous FDs, the event date is specified as the reference search parameter. In the current work, another simple code $-$ non-coincident algorithm $-$ which operates differently is also developed. The two subroutines are used for event identification. One searches for simultaneous events while the second identifies non-simultaneous FDs.

\citep{34} used the coincident code to select simultaneous FDs from seven event catalogs. Each list from the seven CR stations investigated contains a large number of FDs with a minimum and maximum of 238 and 386 FDs respectively. The simple algorithm was recently used to select simultaneous FDs as well as the amplitude of the accompanying diurnal anisotropy \citep[see][]{45}.

Here, the catalogs of FDs selected from INVK and THUL by Prog1 and Prog2 serve as input to the coincident code.

\section{Results and Discussions}

\subsection{FD Event Selection by Prog1}

\begin{center}
	\begin{figure}
	\centering
	\includegraphics[width=10cm]{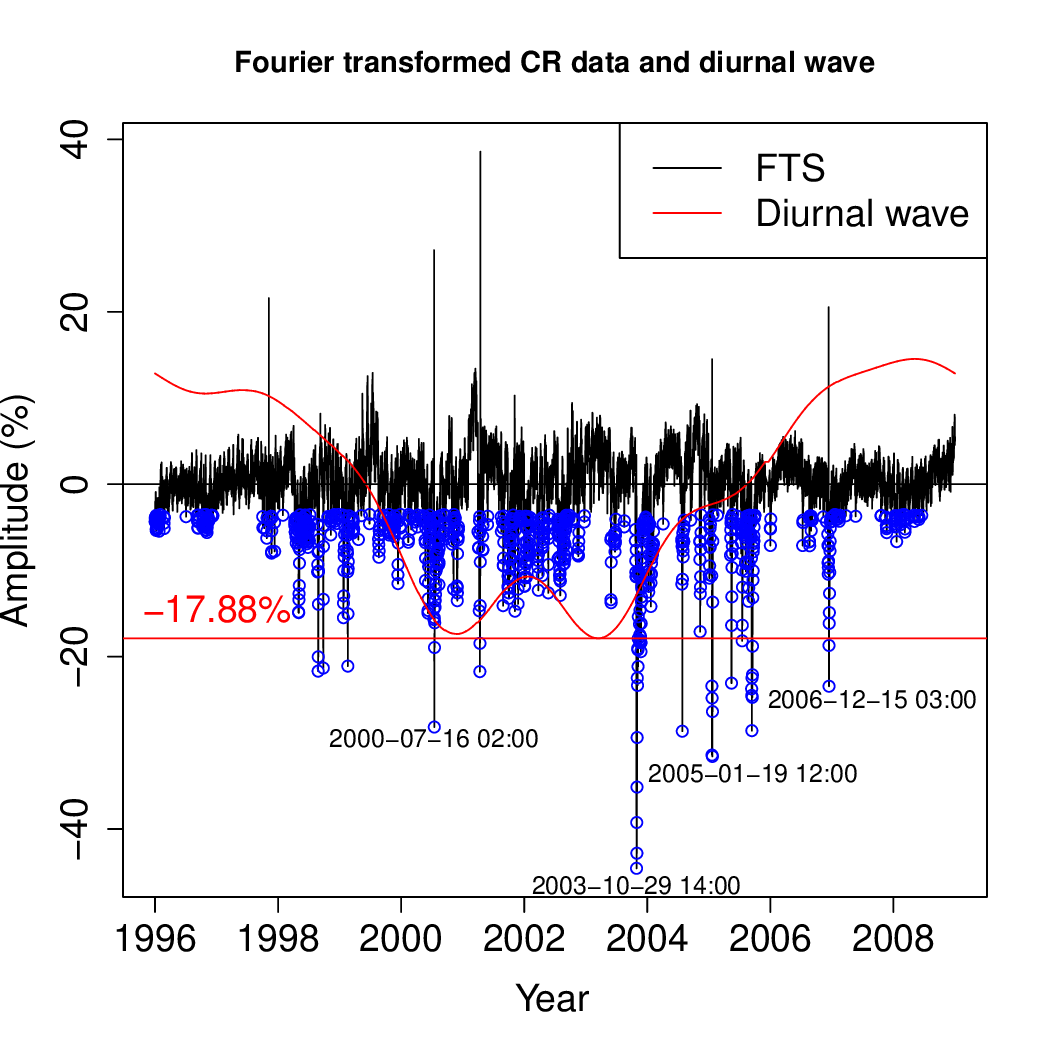}
	\caption{Same as Figures \ref{Figure4} for the Solar Cycle 23.}
	\label{Figure8}
	 \end{figure}
	\end{center}  

\begin{center}
	\begin{figure}
	\centering
	\includegraphics[width=10cm]{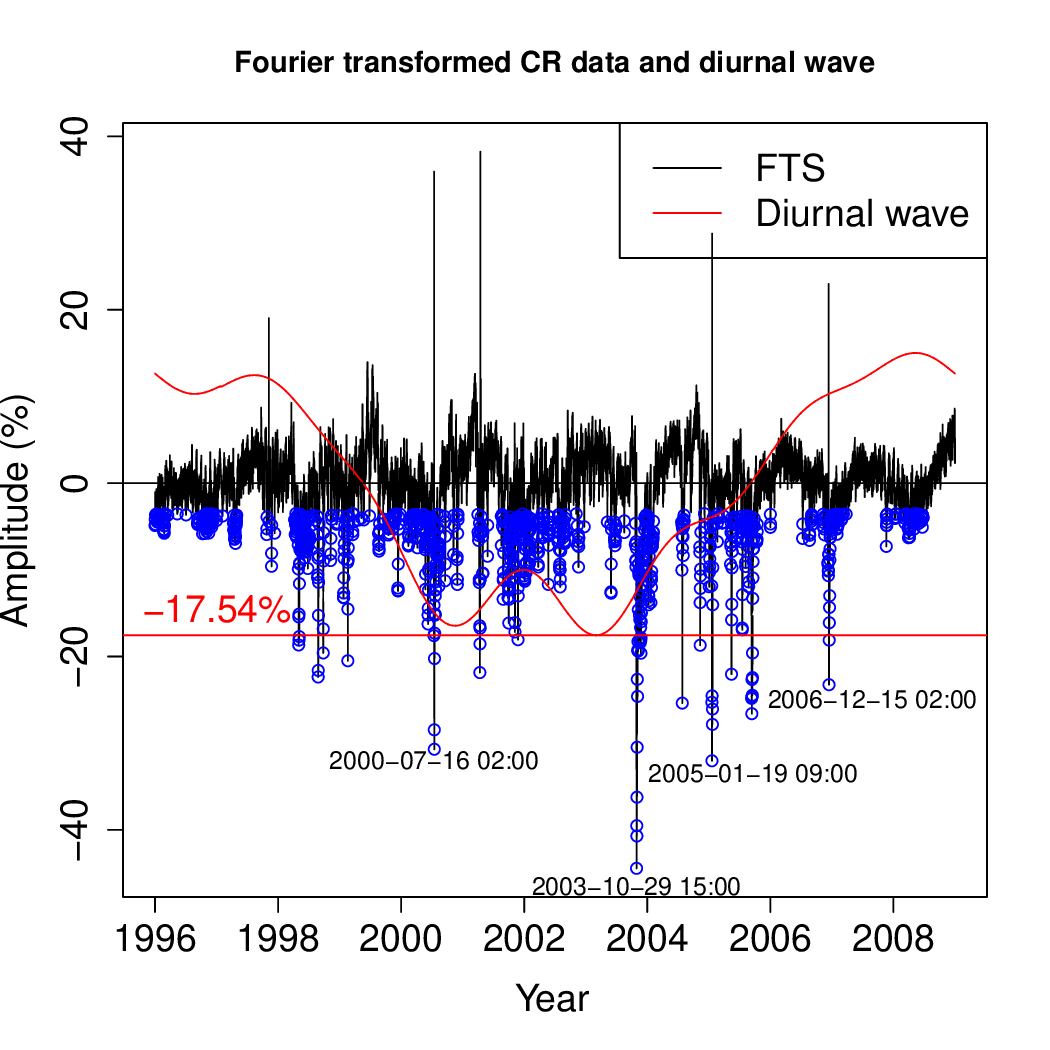}
	\caption{Same as Figures \ref{Figure5} for the Solar Cycle 23. The blue circles are indications of FDs.}
	\label{Figure9}
	 \end{figure}
	\end{center}  

Figures \ref{Figure8} and \ref{Figure9} present the results of FD event selection (Prog1) for Solar Cycle 23 at INVK and THUL stations respectively. The black and red signals are the Fourier-transformed CR data (FTS) and the diurnal wave. The blue circles are reflections of the magnitude and timing of FDs. The raw data and the associated FDs are not reported here.

Some of the large FDs in Figure \ref{Figure3} are also marked in these diagrams for comparative reasons. In the three figures, the series of FDs that happened towards the end of October 2003 is the largest. In Figure \ref{Figure3}, the second to this event in magnitude is the FD of 16/7/2000. However, after removing the 11-year solar cycle effects, Figures \ref{Figure8} and \ref{Figure9} suggest that the event of 19 January 2005 is the second in the rank of FDs that happened within the period. This is followed by the event of 16/7/2000. 

The number of FDs detected at INVK and THUL are respectively 884 and 980. A common threshold of $\mathrm{CR} (\%)\leq -3.5$ \citep[e.g]{46} is used for event selection at the two stations. It should be noted that the number of FDs identified may vary when different normalization baselines are used. With a baseline of $\mathrm{CR} (\%)\leq -3$ and $\mathrm{CR} (\%)\leq -2.5$, for example, \citep{46} respectively identified 3905 and 6797 FDs from CLMX data between 1957-2006.

The amplitudes of the diurnal waves exhibit interesting patterns at the two stations. They tend to follow the 11-year solar cycle, pointing to the periods of high solar activities (e.g. 2000 and 2003). Two peaks of the waves are evident in the diagrams. The two largest amplitudes happened in the years 2001 and 2003. The two years are associated with periods of high solar activity \citep[e.g][]{40}. The year 2001 is the year of solar maximum. While the amplitudes of these waves are nearly equal in Figure \ref{Figure8} (-17.88\%), Figure \ref{Figure9} shows that the amplitude is larger in 2003 (-17.54\%) than in 2001. It is also worth noting that the amplitudes of the waves decreased around 2002 at the two stations.

The amplitudes of the diurnal wave in Figures \ref{Figure8} and \ref{Figure9} reflect those in Figures \ref{Figure4}, \ref{Figure5}, \ref{Figure6} and \ref{Figure7}. Whereas small amplitudes are registered in the year 1996 (Figures \ref{Figure4} and \ref{Figure6}), large amplitudes of the diurnal wave are registered at the two stations in 2003 $-$ year of high solar activity (Figures \ref{Figure6} and \ref{Figure7}). They also reflect the result presented by \citep{34} using daily averaged CR data. 

Figures \ref{Figure8} and \ref{Figure9} also allows the reader to consider FD event simultaneity at the two stations. The exact time is indicated for the four events that are marked in the diagrams. The time differences between the four large events at INVK and THUL stations are in the range of $0 \leq \mathrm{UT} \leq 3$, where UT represents the universal time. In the light of \cite{45} where events are said to be simultaneous if points of minimal CR depressions are observed on the same day (regardless if the differences of the FD minimum at different stations are several hours), the four events might be taken as simultaneous FDs. In the current work, however, only the event of 16/7/2000 whose time difference at the two stations is 0 (zero) is simultaneous at INVK and THUL.

\subsection{Simultaneous/Non-Simultaneous FDs and the Associated Solar/Geophysical Parameters}

Table \ref{table 1} presents the list of 61 large FDs that happened at the same universal time at INVK and THUL stations (List of the low magnitude ($>$ -3\%) events that are simultaneous at the two stations are not presented). Solar wind data (IMF and SW) and D$_{st}$ associated with the FDs are also listed. The Table allows us to comment on the CR intensity modulations at the two stations. Inspection of the first 6 events suggests that THUL measures larger FDs compared with INVK. Nevertheless, the magnitude of the 7th event is larger at INVK than at THUL. On the other hand, the magnitude of the 8th event is nearly equal at the two stations. This points to the varieties of FDs and their different manifestations at different locations \citep{22,27}. This is also reflected in Figure \ref{Figure3}. A close inspection of the diagram shows that INVK sees larger intensity variations for some events. The event of 31/10/2003 is apparently larger at INVK in the diagram. The event is the 44th in Table \ref{table 1}. Its magnitude at INVK and THUL are respectively -42.80\% and -40.71\%. Another event that seems to be a simultaneous FD at the two stations in Figure \ref{Figure3} is the event of 16/7/2000. The longer tail of the event at THUL suggests that its magnitude is larger than at INVK. This difference is also reflected in Table \ref{table 1}. The magnitude of the events at INVK and THUL are respectively -28.17 and -30.70\%.

The mean, maximum, minimum, the 25 and 75\% of the FDs at INVK are 10.42 $\pm$ 0.90\%, 42.81\%, 3.89\%, 5.69 (25\%, 25 percentile) and 13.06 (75\%, 75 percentile). For THUL stations, the mean, maximum, minimum, 25, and 75\% of the FDs are 10.31 $\pm$ 0.88\%, 40.71\%, 3.73\%, 6.05 (25\%) and 12.54 (75\%). 

Given the large number of FDs detected at INVK (884) and THUL (980), one may infer from the result presented in Table \ref{table 1} that non-simultaneous FDs are much more in number than events that happen simultaneously at the two stations. While Table \ref{table 1} reports that only 7 FDs are simultaneous in the year 2003, table 3 of \cite{45} shows that simultaneous FDs are more in number, depending on pairs of stations of interest. The number of simultaneous FDs at Sanae (SNAE) and Moscow (MOSC) stations is, for instance, 16. For South Pole (SOPO) and ClMX stations, 33 events are simultaneously detected. \cite{45} attributes the large number difference between simultaneous FDs selected with daily and hourly data to the speculated increase in diurnal anisotropy \citep[see][and the references therein]{2,47,30} in hourly averaged CR data. The phase shift \citep{48} arising from CR anisotropy impacts significantly on the global simultaneity of FDs. Additionally, \citep{38} speculate that binning artifact may play a significant role in the outcome of the simultaneity test between two temporal resolutions.

The number of FDs at INVK which are not simultaneously detected at THUL station is 823 while 919 events (result not presented), at THUL, do not happen at the same universal time at INVK. Although one could arrive at these numbers by simple arithmetic, a subroutine of the coincident algorithm, capable of isolating non-simultaneous FDs from two event catalogs is required to separate the 61 simultaneous events from the large number of FDs at each station.

The mean, maximum, minimum, the 25 and 75\% of the FDs at INVK are 7.11 $\pm$ 0.16\%, 44.56\%, 3.50\%, 4.23 (25\%, 25 percentile) and 8.04 (75\%, 75 percentile). For THUL stations, the mean, maximum, minimum, 25, and 75\% of the FDs are 7.09 $\pm$ 0.15\%, 44.44\%, 3.50\%, 4.27 (25\%) and 7.84 (75\%). A comparison of these statistics with those of the simultaneous FDs suggests that the magnitude of the simultaneous FDs is, on average, larger than those of the non-simultaneous events. This agrees with the indications of \cite{25} and \citep{27}. The time difference between the largest event at INVK (44.56\%) and THUL (44.44\%) is an hour (see Figures \ref{Figure8} and \ref{Figure9}).

\begin{table}
\caption{List of large and simultaneous FDs at INVK and THUL CR stations and the associated geomagnetic storm index (D$_{st}$) solar wind data .}
\label{table 1}
\centering
\begin{tabular}{rlrrrrrr}
  \hline
Order & Year & Hour & INVK (\%) & THUL (\%) & IMF (nT) & SW (Km/s) & D$_{st}$ (nT) \\ 
  \hline
1 & 1996-10-09 &  14 & -4.95 & -5.84 & 4.70 & 494 & -26 \\ 
  2 & 1998-05-10 &   07 & -4.74 & -7.68 & 6.30 & 481 & -46 \\ 
  3 & 1998-06-14 &   05 & -4.90 & -6.90 & 11.40 & 348 & -10 \\ 
  4 & 1998-08-24 &  10 & -9.32 & -10.57 & 3.50 & 423 & -20 \\ 
  5 & 1998-08-28 &   02 & -13.93 & -15.26 & 6.50 & 589 & -84 \\ 
  6 & 1998-11-10 &   00 & -4.38 & -6.37 & 11.30 & 432 & -93 \\ 
  7 & 1999-01-23 &  13 & -15.48 & -13.17 & 13.10 & 578 &  -9 \\ 
  8 & 1999-08-23 &   09 & -7.74 & -7.59 & 9.40 & 402 & -48 \\ 
  9 & 1999-09-16 &   02 & -5.26 & -4.14 & 6.20 & 548 & -18 \\ 
  10 & 1999-10-26 &  10 & -4.81 & -4.19 & 5.20 & 470 & -24 \\ 
  11 & 1999-12-12 &  20 & -10.51 & -12.09 & 8.10 & 657 & -17 \\ 
  12 & 1999-12-14 &   00 & -9.32 & -12.37 & 12.00 & 442 & -33 \\ 
  13 & 2000-03-01 &   07 & -5.25 & -6.18 & 9.00 & 518 & -20 \\ 
  14 & 2000-05-22 &   09 & -6.47 & -7.39 & 6.40 & 445 &   4 \\ 
  15 & 2000-06-10 &   06 & -13.29 & -14.43 & 8.10 & 546 & -44 \\ 
  16 & 2000-06-24 &   06 & -5.08 & -6.20 & 12.90 & 574 & -31 \\ 
  17 & 2000-07-16 &   02 & -28.17 & -30.70 & 45.90 & 970 & -278 \\ 
  18 & 2000-07-19 &   00 & -13.54 & -15.08 & 3.90 & 571 & -33 \\ 
  19 & 2000-08-13 &   02 & -9.09 & -7.81 & 19.80 & 554 & -72 \\ 
  20 & 2000-11-08 &   04 & -6.63 & -3.73 & 16.00 & 445 & -34 \\ 
  21 & 2001-04-04 &  20 & -7.41 & -6.05 & 13.40 & 713 & -14 \\ 
  22 & 2001-04-12 &   09 & -21.74 & -21.85 & 19.60 & 645 & -134 \\ 
  23 & 2001-08-31 &   02 & -8.56 & -10.26 & 6.60 & 412 &  -9 \\ 
  24 & 2001-10-06 &   02 & -6.56 & -5.92 & 3.40 & 421 & -14 \\ 
  25 & 2001-11-24 &  18 & -12.17 & -12.11 & 22.00 & 827 & -205 \\ 
  26 & 2001-11-26 &   01 & -9.37 & -13.01 & 4.60 & 638 & -86 \\ 
  27 & 2002-01-04 &   00 & -10.79 & -11.04 & 5.70 & 320 &  -7 \\ 
  28 & 2002-01-07 &   00 & -6.72 & -6.59 & 7.60 & 330 &   2 \\ 
  29 & 2002-01-10 &  18 & -7.87 & -4.52 & 18.60 & 581 & -37 \\ 
  30 & 2002-01-13 &  10 & -9.23 & -9.29 & 5.10 & 528 & -24 \\ 

  31 & 2002-01-17 &   05 & -5.19 & -4.83 & 5.40 & 358 &  -9 \\ 
  32 & 2002-01-19 &  11 & -4.98 & -5.71 & 13.00 & 372 &  24 \\ 
  33 & 2002-03-20 &  17 & -8.41 & -8.00 & 17.20 & 520 &  11 \\ 
  34 & 2002-03-24 &  16 & -10.37 & -9.39 & 14.10 & 441 & -94 \\ 
  35 & 2002-04-18 &   09 & -10.59 & -5.20 & 13.80 & 492 & -111 \\ 
  36 & 2002-04-20 &   03 & -7.39 & -6.29 & 18.60 & 563 & -93 \\ 
  37 & 2002-07-26 &  14 & -5.69 & -5.57 & 7.20 & 429 & -14 \\ 
  38 & 2002-09-08 &   02 & -7.63 & -5.64 & 19.30 & 495 & -152 \\ 
  39 & 2002-11-21 &   06 & -5.60 & -6.31 & 24.50 & 531 & -79 \\ 
  40 & 2003-05-30 &  13 & -13.40 & -12.54 & 16.50 & 573 & -63 \\ 
  41 & 2003-08-18 &   04 & -4.25 & -4.35 & 19.70 & 492 & -53 \\ 
  42 & 2003-10-23 &   01 & -5.25 & -4.65 & 9.90 & 524 & -13 \\ 
  43 & 2003-10-28 &  10 & -10.70 & -7.50 & 12.90 & 759 & -32 \\ 
  44 & 2003-10-31 &  11 & -42.80 & -40.71 & 14.40 & 1003 & -55 \\ 
  45 & 2003-11-20 &  15 & -18.12 & -18.41 & 55.80 & 614 & -171 \\ 
  46 & 2003-11-28 &   01 & -13.08 & -14.03 & 5.20 & 418 &  -6 \\ 
  47 & 2004-02-03 &   08 & -7.05 & -6.15 & 6.40 & 614 & -34 \\ 
  48 & 2004-11-07 &  21 & -6.66 & -7.35 & 45.10 & 660 & -74 \\ 
  49 & 2005-01-06 &   02 & -7.25 & -8.75 & 6.80 & 568 & -11 \\ 
  50 & 2005-01-21 &  22 & -31.58 & -27.83 & 22.40 & 901 & -89 \\ 
  51 & 2005-05-14 &   01 & -3.89 & -4.09 & 4.60 & 525 & -18 \\ 
  52 & 2005-05-15 &   07 & -23.07 & -22.04 & 54.30 & 922 & -229 \\ 
  53 & 2005-05-16 &   03 & -16.37 & -15.78 & 14.10 & 704 & -99 \\ 
  54 & 2005-05-17 &   03 & -13.06 & -13.97 & 7.00 & 572 & -68 \\ 
  55 & 2005-06-16 &  10 & -4.58 & -4.05 & 18.00 & 538 & -22 \\ 
  56 & 2005-08-06 &  22 & -9.21 & -9.14 & 8.80 & 595 & -29 \\ 
  57 & 2005-08-08 &   04 & -10.10 & -6.65 & 4.70 & 571 & -15 \\ 
  58 & 2005-08-24 &  17 & -13.58 & -11.94 & 23.70 & 657 & -133 \\ 
  59 & 2005-09-22 &  13 & -7.36 & -7.57 & 5.50 & 326 & -19 \\ 
  60 & 2006-12-10 &   09 & -10.44 & -10.15 & 4.90 & 625 & -24 \\ 
  61 & 2006-12-17 &   01 & -14.90 & -16.11 & 6.70 & 650 & -36 \\ 
   \hline
\end{tabular}
\end{table}

\begin{table}
\caption{Correlation results of FDs and the associated solar and geomagnetic data. The symbols on the diagrams are as follows: "r" represents the Pearson's product-moment correlation coefficient, N represents the number of FDs in a group, "ALL" denote all the non-simultaneous FDs at INVK, THUL, and for the IZMIRAN catalogue, "t" for the t static, "R${^2}$" is the coefficient of determination, FD\_invk and FD\_thul stand for FDs at INVK and THUL stations respectively, and "none" indicate relations that are not significant at any level.}
\label{table 2}
\centering
\begin{tabular}{rlllll}
  \hline
Order & Variables & r & t & R${^2}$ & Sig level (\%) \\ 
  \hline
$-$ & Simultaneous FDs (N = 61)&  &  &  &  \\ 
  1 & FD\_invk/FD\_thul & 0.97 & 30.07 & 0.94 & 99.8 \\ 
  2 & FD\_invk/IMF & 0.40 & 3.32 & 0.16 & 99.8 \\ 
  3 & FD\_thul/IMF & 0.37 & 3.09 & 0.14 & 99 \\ 
  4 & FD\_invk/SW & 0.74 & 8.56 & 0.55 & 99.8 \\ 
  5 & FD\_thul/SW & 0.71 & 7.65 & 0.50 & 99.8 \\ 
  6 & FD\_ink/DST & -0.50 & 4.41 & 0.25 & 99.8 \\ 
  7 & FD\_thul/DST & -0.50 & 4.38 & 0.25 & 99.8 \\ 
  8 & IMF/DST & -0.74 & 8.45 & 0.55 & 99.8 \\ 
  9 & SW/DST & -0.60 & 5.75 & 0.40 & 99.8 \\ 
  10 & SW/IMF & 0.52 & 4.67 & 0.27 & 99.8 \\ 
  $-$ & Non-Simultaneous (ALL, INVK, N = 823) &  &  &  &  \\ 
  1 & FD\_invk/IMF & 0.0 & 0.0 & 0.0 & none \\ 
  2 & FD\_invk/SW & 0.26 & 7.91 & 0.07 & 99.8 \\ 
  3 & FD\_invk/DST & -0.39 & 12.13 & 0.15 & 99.8 \\ 
  $-$ & Group A (INVK, N = 478) &  &  &  &  \\ 
  1 & FD\_invk/IMF & 0.0 & 0.1 & 0.0 & none \\ 
  2 & FD\_invk/SW & 0.36 & 8.33 & 0.13 & 99.8 \\ 
  2 & FD\_invk/DST & -0.34 & 7.93 & 0.12 & 99.8 \\ 
  $-$ & Group B (INVK, N = 345) &  &  &  &  \\ 
  1 & FD\_invk/IMF & -0.04 & 0.71 & 0.0 & 50 \\ 
  2 & FD\_invk/SW & 0.02 & 0.43 & 0.0 & none \\ 
  3 & FD\_invk/DST & -0.01 & 0.22 & 0.0 & none \\ 
  $-$ & Non-Simultaneous (ALL, THUL, N = 919) &  &  &  &  \\
  1 & FD\_thul/IMF & -0.01 & 0.40 & 0.00 & none \\ 
  2 & FD\_thul/SW & 0.22 & 6.74 & 0.05 & 99.8 \\ 
  3 & FD\_thul/DST & -0.45 & 15.99 & 0.22 & 99.8 \\ 
  $-$ & Group A (THUL, N = 527) &  &  &  &  \\ 
  1 & FD\_thul/IMF & 0.33 & 7.91 & 0.11 & 99.8 \\ 
  2 & FD\_thul/SW & 0.43 & 10.84 & 0.18 & 99.8 \\ 

  3 & FD\_thul/DST & -0.45 & 11.65 & 0.21 & 99.8 \\ 
  $-$ & Group B (THUL, N = 392)&  & &  &  \\ 
  1 & FD\_thul/IMF & 0.03 & 0.66 & 0.0 & none \\ 
  2 & FD\_thul/SW & 0.02 & 0.32 & 0.0 & none \\ 
  3 & FD\_thul/DST & 0.0 & 0.0 & 0.0 & none \\ 
  $-$ & IZMIRAN (ALL FDs, N = 1476) &  &  &  &  \\ 
  1 & FD/IMF & 0.0 & 0.05 & 0.0 & none \\ 
  2 & FD/SW & 0.20 & 7.88 & 0.04 & 99.8 \\ 
  3 & FD/DST & -0.05 & 2.01 & 0.003 & 95 \\ 
  $-$ & Group A (N = 134) &  &  &  &  \\ 
  1 & FD/IMF & 0.44 & 5.67 & 0.0 & 99.8 \\ 
  2 & FD/SW & 0.48 & 6.34 & 0.04 & 99.8 \\ 
  3 & FD/DST & -0.09 & 1.07 & 0.003 & 50 \\ 
  $-$ & Group B (N = 1342) &  &  &  &  \\ 
  1 & FD/IMF & -0.01 & 0.37 & 0.0 & none \\ 
  2 & FD/SW & -0.02 & 0.58 & 0.04 & none \\ 
  3 & FD/DST & -0.04 & 1.28 & 0.003 & 80 \\ 
   \hline
\end{tabular}
\end{table}

\section{Summary}
Although articles investigating Forbush effects \citep{22} started appearing in the literature in the 1930s \citep{49}, the selection of FDs from a given NM data is one of the difficult aspects of the subject. This is due to the ever-present CR diurnal anisotropies. Though the bias implications of the periodic and the enhanced diurnal anisotropy on timing and magnitude as well as on the number of FDs detected in a given period/data has long been recognized \citep[e.g][]{2,4,48,50}, several of the existing FD lists are prepared without adjusting for CR anisotropy. This could explain the conflicting and misleading conclusions in some past articles utilizing different FD catalogs \citep[see][for discussion]{10,31,33,46}.

Graphical illustration of FD global simultaneity was indeed attempted by some researchers in the early 1970's \citep[e.g][]{2}, \cite{25} asserted that they were the first to investigate the global simultaneous/non-simultaneous FDs as well as the associated solar wind data. Besides the Oh et al. research team \citep[see also][]{26,29,51} and the contribution from \cite{27}, \cite{45}, and \cite{38}, we are not aware of other publications that are dedicated to the analysis of global simultaneity of FDs. 

While other works that investigate the characteristics of simultaneous and non-simultaneous FDs at different points on Earth failed to take the impact of anisotropy on FD profiles into consideration, the recent article of \cite{45} carried out a detailed analysis of CR anisotropy to examine the properties of the simultaneous and non-simultaneous FDs. The current work is a follow-up on the new approach.

Some of the main differences between this and \cite{45} lie in the differences in the data resolution and period. While \cite{45} employed daily averaged data over ten CR stations for one year, the current work uses hourly averaged CR data over two NMs for a full solar cycle. Since this is a preliminary effort to investigate FD global event simultaneity using FDs automatically selected from hourly averaged data, result validation was of topmost priority.

A close inspection of Table \ref{table 1} and the associated correlation results in Table \ref{table 2} suggests that the FDs identified by the computer code might be related to solar as well as geophysical data. Although there is no one-to-one relationship between FDs and D$_{st}$, for example, some of the large FDs in the Table are linked with large negative D$_{st}$. A detailed investigation of solar wind conditions generating such events might be taken up as case studies.
The present analysis may be repeated using FDs selected by Prog2. All the events presented and discussed in the result section are identified by Prog1.
\section*{Acknowledgements}
 Huge thanks are due to those maintaining the following websites:\\ \url{http://cr0.izmiran.ru/common},\\ \url{http://spaceweather.izmiran.ru/eng/dbs.html} \\ and \\ \url{https://omniweb.gsfc.nasa.gov/html/ow_data.html}.
\pagebreak


\label{lastpage}

\end{document}